\documentclass{pasj00}

\begin{document}
\SetRunningHead{Kajisawa et al.}{MOIRCS Deep Survey. X.}

\title{MOIRCS Deep Survey. X. Evolution of Quiescent Galaxies
as a Function of Stellar Mass at $0.5<z<2.5$}

\author{Masaru \textsc{Kajisawa}\altaffilmark{1}, 
Takashi \textsc{Ichikawa}\altaffilmark{2}, 
Tomohiro \textsc{Yoshikawa}\altaffilmark{3},
Toru \textsc{Yamada}\altaffilmark{2}, 
Masato \textsc{Onodera}\altaffilmark{4},\\ 
Masayuki \textsc{Akiyama}\altaffilmark{2}, 
Ichi \textsc{Tanaka}\altaffilmark{5}}

\email{kajisawa@cosmos.ehime-u.ac.jp}

\altaffiltext{1}{Research Center for Space and Cosmic Evolution, Ehime University, Bunkyo-cho 2-5, Matsuyama 790-8577, Japan}
\altaffiltext{2}{Astronomical Institute, Tohoku University, Aramaki,
Aoba, Sendai 980--8578, Japan}
\altaffiltext{3}{Koyama Astronomical Observatory, Kyoto Sangyo University, Motoyama, Kamigamo, Kita-ku, Kyoto 603--8555, Japan}
\altaffiltext{4}{Institute for Astronomy, ETH Zurich, Wolfgang-Pauli-strasse 27, 8093 
Zurich, Switzerland} 
\altaffiltext{5}{Subaru Telescope, National Astronomical Observatory
of Japan, 650 North Aohoku Place, Hilo, HI 96720, USA}


%

\KeyWords{galaxies:evolution --- galaxies:formation --- galaxies:high-redshift} 

\maketitle

\begin{abstract}
We study the evolution of quiescent galaxies at $0.5<z<2.5$ as a function of 
stellar mass, using very deep NIR imaging data taken with the Multi-Object Infrared 
Camera and Spectrograph on the Subaru Telescope in the 
GOODS-North region.
The deep NIR data allow us to construct a stellar mass-limited sample of 
quiescent galaxies down to $\sim 10^{10}$ M$_{\odot}$ even at $z\sim2$ for the 
first time.
We selected quiescent galaxies with age/$\tau>6$ by  
performing SED fitting of 
 the multi broad-band photometry from the $U$ to $Spitzer$ 
5.8$\mu$m bands with the population synthesis model of \citet{bru03} where
exponentially decaying star formation histories are assumed. 
The number density of quiescent galaxies 
increases by a factor of $\sim3$ 
from $1.0<z<1.5$ to $0.5<z<1.0$, and by a factor of $\sim10$ from 
$1.5<z<2.5$ to $0.5<z<1.0$, while that of star-forming galaxies with age/$\tau<4$ 
increases only by factors of $\sim2$ and $\sim3$ in the same redshift ranges.
At $0.5<z<2.5$, 
the low-mass slope of the stellar mass function of quiescent galaxies is 
$\alpha \sim$ 0 -- 0.6, which is significantly flatter than those of star-forming galaxies
 ($\alpha \sim$ -1.3 -- -1.5). 
As a result, 
the fraction of quiescent galaxies in the overall galaxy population 
increases with stellar mass in the redshift range. 
The fraction of quiescent galaxies at 10$^{11}$--10$^{11.5}$ M$_{\odot}$ 
increases from $\sim$20--30\% at 
$z\sim2$ to $\sim$40--60\% at $z\sim0.75$, while 
that at 10$^{10}$--10$^{10.5}$ M$_{\odot}$ 
 increases from $\lesssim$ 5\% to $\sim$15\% in the same redshift range. 
These results could suggest that  the quenching of star formation had been 
more effective in more massive galaxies at $1\lesssim z \lesssim 2$.  
Such a mass-dependent quenching could explain 
 the rapid increase of the number density of $\sim$ M$^*$ 
galaxies relative to lower-mass galaxies at $z\gtrsim$ 1--1.5.

\end{abstract}

\section{Introduction}
Determining how  stars have been formed in galaxies is crucial for 
understanding galaxy formation and evolution. 
In the present universe, it is known that galaxies can be well separated into 
two populations, namely, passive-evolving galaxies with red colors 
and star-forming galaxies with blue colors (e.g., \cite{kau03}; \cite{bal04}; 
\cite{bri04}). 
Bimodal color distribution which consists of these two populations
has also been observed up to $z\sim$ 1--2 (e.g., \cite{bel04}; \cite{wei05}; 
\cite{fra07}; \cite{cir07}; \cite{cas08}; \cite{wil09}; \cite{bra09}).
At $z<1$, several studies have found that the stellar mass density of passively-evolving 
galaxies increases by a factor of $\sim 2$ from $z\sim1$ to $z\sim0$, while 
that of star-forming galaxies evolves only very mildly (e.g., \cite{bor06};  
\cite{fab07}; \cite{poz09}; 
\cite{ilb10}).  Since passively-evolving galaxies hardly accumulate stellar mass 
only by star formation, these results suggest that some fraction of 
star-forming galaxies migrates into the quiescent population by quenching of 
star formation (\cite{bel04}; \cite{bel07}; \cite{fab07}). 
Since recent studies suggest that the star formation rates in star-forming galaxies 
seem to be simply almost proportional to their stellar mass at least at $z\lesssim 2$ 
(e.g., \cite{elb07}; \cite{dad07}; \cite{pan09}; \cite{kaj10}), 
how the quenching of  star formation has occurred in galaxies becomes a key issue to 
understand the various star formation histories of galaxies. 

On the other hand, it is also known from studies of the `fossil record' of 
present-day galaxies that the star formation histories of galaxies seem to 
 depend strongly on their stellar mass (e.g., \cite{hea04}; \cite{jim05}).
In the local universe, massive galaxies tend to have redder color and older stellar 
population, while most low-mass galaxies are young and actively star-forming 
(\cite{kau03}; \cite{bri04}). 
Since the pioneering work by \citet{cow96},  
many observational studies at higher redshift have suggested that more massive galaxies 
have older stellar population even at $z\sim1$ and  formed 
their stars earlier and more rapidly than low-mass galaxies (e.g., \cite{bri00}; 
\cite{jun05}; \cite{feu05}; \cite{bun06}; \cite{ver08}).   
For quiescent galaxies with little star formation, 
\citet{ilb10} and \citet{poz09} reported that the number density of 
low-mass quiescent galaxies increases more rapidly 
than massive galaxies from $z\sim1$ to $z\sim0$. 
Recently, \citet{pen10} proposed that the mass-dependent quenching mechanism 
which ceases star formation preferentially for more massive galaxies is needed 
to explain no evolution of the characteristic mass M$^*$ for star-forming galaxies. 
On the other hand, the increase of low-mass quiescent galaxies at $z\lesssim 1$ is 
explained by the environmental effects in their scenario.  
Since the clustering of both bright and faint quiescent galaxies has been 
 observed to be stronger than star-forming galaxies (e.g., \cite{zah05}; \cite{mac08}), 
massive dark matter halos may play some role in the quenching of star formation. 
It is important to investigate how the evolution of quiescent galaxies and 
the quenching of star formation depend on stellar mass at higher redshift, 
as previous studies of the evolution of the global star formation rate density and 
stellar mass density in the universe suggest that the cosmic star formation rate 
has a peak around $z\sim2$ and that a significant fraction of the stellar mass in 
the present universe had been formed at $z\sim$ 1--3 (e.g., \cite{hop06}; \cite{wil08}). 

At $z\sim$ 1.5--2, several studies have found massive quiescent galaxies 
by spectroscopic observations of red objects (e.g., \cite{cim04}; \cite{sar05}; 
\cite{abr07}; 
\cite{kri06}; \cite{kri08}; \cite{kri09}; \cite{ono10}) and by color selection techniques 
 with imaging data (e.g., \cite{dad04}; \cite{wil09}; \cite{ilb10}; \cite{cam10}) 
or SED fitting techniques with multi-band photometry
 (e.g., \cite{gra07}; \cite{fon09}; \cite{whi10}). 
\citet{ilb10} reported that the number density of massive quiescent galaxies 
with M$_{\rm star} \sim 10^{11}$ M$_{\odot}$ rapidly increases from $z\sim 1.75$ 
and to $z\sim1$, while it evolves only mildly at $z\lesssim 1$.
Furthermore, 
\citet{fon09} also found that the quiescent fraction in massive galaxies with M$_{\rm star} 
> 7\times 10^{10}$ M$_{\odot}$ significantly increases between $z\sim2$ and $z\sim1$. 
The low fraction of quiescent galaxies at $z\gtrsim2$ reflects that there are 
many active star-forming galaxies which dominates the massive galaxy population 
(e.g., \cite{pap07}; \cite{gra07}).   
Thus $1<z<2$ seems to be the formation epoch of  
these massive quiescent galaxies.  

On the other hand, \citet{kaj09} found that the number density of 
$\sim $ M$^*$ ($\sim 10^{11}$ M$_{\odot}$) galaxies evolves more strongly 
than low-mass galaxies at $1\lesssim z \lesssim 3$. 
Such mass-dependent evolution of the number density may indicates that 
the star formation history and/or stellar mass assembly of galaxies already 
strongly depended on stellar mass at that epoch when most stars seen in the 
present universe were formed. 
Therefore it is interesting to investigate the mass-dependence of the evolution of 
quiescent galaxies up to $z\sim2$. 
However, low-mass quiescent galaxies at $z\gtrsim$ 1--1.5 has not yet been investigated 
so far.  Since quiescent galaxies with relatively old stellar population tend to have high 
stellar mass-to-luminosity (M/L) ratios, the observed fluxes of quiescent 
galaxies with a given stellar mass become relatively faint (e.g., \cite{fon03}). 
In order to sample low-mass quiescent galaxies with high completeness, 
very deep near-infrared (NIR) data are required (e.g., \cite{kaj06b}). 

In this paper, we study the evolution of quiescent galaxies at $0.5<z<2.5$ 
as a function of stellar mass using very deep NIR data  
 from MOIRCS Deep Survey (MODS, \cite{kaj06}; \cite{ich07}). 
The MODS data reach $\sim$ 23--24 Vega magnitude ($\sim$ 25--26 AB magnitude) 
in the $K_{s}$ band, and they allow us to construct a stellar mass 
limited sample of quiescent galaxies 
down to $\sim 10^{10}$ M$_{\odot}$ even at $z\sim2$.
Section 2 describes the observational data and the procedures of 
source detection and photometry.
Details of the SED fitting analysis of the detected objects are given in Section 3.
In Section 4, we use the results of the SED fitting to select quiescent galaxies 
and construct a stellar mass-limited quiescent sample. 
We show the evolution of the number density and fraction of quiescent galaxies 
as a function of stellar mass in Section 5. In Section 6, we 
compare the results with previous studies and discuss their implications.  
A summary is presented in Section 7. 

We use a cosmology with H$_{\rm 0}$=70 km s$^{-1}$ Mpc$^{-1}$, 
$\Omega_{\rm m}=0.3$ and $\Omega_{\rm \Lambda}=0.7$.
The Vega-referred magnitude system is used throughout this paper, 
unless stated otherwise.



\section{Observational Data and Photometry}
\label{sec:obs}
We use the $K_{s}$-selected sample of the MODS in the GOODS-North 
region (\cite{kaj09}, hereafter K09), 
which is based on our deep $JHK_{s}$-bands 
imaging data taken with MOIRCS \citep{suz08} on the Subaru telescope.
Four MOIRCS pointings cover $\sim70$\% of the GOODS-North region ($\sim$ 103.3
arcmin$^2$, hereafter referred as ``wide'' field) 
and the data reach $J=24.2$, $H=23.1$, $K=23.1$ (5$\sigma$, 
Vega magnitude). One of the four pointings 
is the ultra-deep field of the MODS ($\sim$ 28.2 arcmin$^2$, hereafter ``deep''
 field),  
where the data reach $J=25.1$, $H=23.7$, $K=24.1$. A full description of the 
observations, reduction, and quality of the data is presented in a separate 
paper \citep{kaj11}.

The source detection was performed in the $K_{s}$-band image using the SExtractor
image analysis package \citep{ber96}. At first, we limited the samples 
to $K_{s}<23$ and $K_{s}<24$ for the wide and deep fields, where the detection 
completeness for point sources is more than 90\% \citep{kaj11}. 
Then we measured 
the optical-to-MIR SEDs of the sample objects, using the publicly available
multi-wavelength data in the GOODS field, namely KPNO/MOSAIC ($U$ band, 
\cite{cap04}), $Hubble$ $Space$ $Telescope$/Advanced Camera for Surveys 
($HST$/ACS; $B$, $V$, $i$, $z$ bands, version 2.0 data; M. Giavalisco et al. 
2010, in preparation; \cite{gia04}) and $Spitzer$/IRAC (3.6$\mu$m, 4.5$\mu$m,
 5.8$\mu$m, DR1 and DR2; M. Dickinson et al. 2010, in preparation), as well as 
the MOIRCS $J$- and $H$-bands images. Details of the multi-band aperture 
photometry are presented in \citet{kaj11}. Following K09, 
we used objects which are detected above 
2$\sigma$ level in more than two other bands in addition to the 5$\sigma$ detection
in the $K_{s}$-band,  
because it is difficult to estimate the photometric redshift and stellar mass of 
those detected only in one or two bands. The number of those excluded by 
this criterion is negligible (21/6402 and 42/3203 for the wide and deep fields, 
respectively).

We also used public $Spitzer$/MIPS 24$\mu$m data in 
the GOODS-North (DR1+, M. Dickinson et al., in preparation) 
to measure 24$\mu$m fluxes of the sample galaxies.
The IRAF/DAOPHOT package \citep{ste87} was used for the photometry 
to deal with many blended sources on the image properly. 
The DAOPHOT software fitted blended sources in a crowded region simultaneously 
with the PSF of the image. 
We used the source positions in the MOIRCS $K_{s}$-band image 
as a prior for the centers of the fitted PSFs in the photometry.
For most objects, for which the residual of the fitting is negligible, 
the 5$\sigma$ limiting flux is $\sim$ 20 $\mu$Jy. 
Details of the photometry on the MIPS image and the error estimate are 
described in \citet{kaj10} and \citet{kaj11}.

\section{SED fitting analysis}
\label{sec:analysis}
K09 estimated the redshift and stellar mass of 
the $K_{s}$-selected galaxies mentioned above and 
constructed a stellar mass-limited sample to study the evolution of 
the stellar mass function. 
We here use the same photometric redshift as in K09 and 
carry out a similar SED fitting of the multi-band photometry 
where the grid of model parameters is slightly changed from that in K09 
in order to estimate age and star-formation time scale $\tau$ (see below) 
optimally for the selection of  quiescent galaxies.

In K09, we show the photometric redshifts agree well with spectroscopic 
redshifts (Figure 1 in K09). If available, we adopted spectroscopic redshifts
from the literature (\cite{coh00}; \cite{coh01}; \cite{daw01}; 
\cite{wir04}; \cite{cow04}; \cite{tre05};  
\cite{red06}; \cite{bar08}; \cite{yos10}), and performed the 
SED fitting fixing the redshift to each spectroscopic value for these galaxies.

We performed the SED fitting of the multi-band photometry
($UBVizJHK$, 3.6$\mu$m, 4.5$\mu$m, and 5.8$\mu$m) with 
a population synthesis model.  
We adopted the GALAXEV population synthesis model \citep{bru03} for 
 direct comparison with previous studies (e.g., \cite{gra07}; \cite{fon09}).
In the model, we assumed exponentially decaying  star formation histories
with the decaying timescale $\tau$ ranging between 0.1 and 100 Gyr.
We used Calzetti extinction law \citep{cal00} in the range of $E(B-V)=$ 0.0-1.0. 
Metallicity is changed from 1/50 to 1 solar metallicity. 
The model age is changed from 50 Myr 
to the age of the universe at the observed redshifts. 
We used 18 grid of $\tau$ and 60 grid of age (smaller number of grid at higher redshift 
depending on the age of the universe), which are roughly equally spaced on a 
logarithmic scale and are slightly finer than those used in 
K09 to select quiescent galaxies by a criterion with age/$\tau$ (see the next section). 
We assume Salpeter IMF \citep{sal55} with lower and upper 
mass limits of 0.1 and 100 M$_{\odot}$ for easy comparison with the results
in other studies.
If we assume the Chabrier-like IMF \citep{cha03}, the stellar mass is reduced 
by a factor of $\sim$ 1.8. 
The redshifts mentioned above and stellar M/L ratios from 
the best-fit templates were used to calculate the stellar mass.
The estimated stellar mass is reasonably consistent with that in K09,  
and the slight change in the grid of the model parameters does not affect the estimate 
of the stellar mass. 
 The uncertainty of the stellar mass  
is estimated taking into account of the photometric redshift error for those 
without spectroscopic redshift,  and is discussed in detail in K09.
In the next section, we will discuss the effect of the stellar mass error on the sample 
selection.

\begin{figure*}[t]
  \begin{center} 
    \FigureFile(170mm,160mm){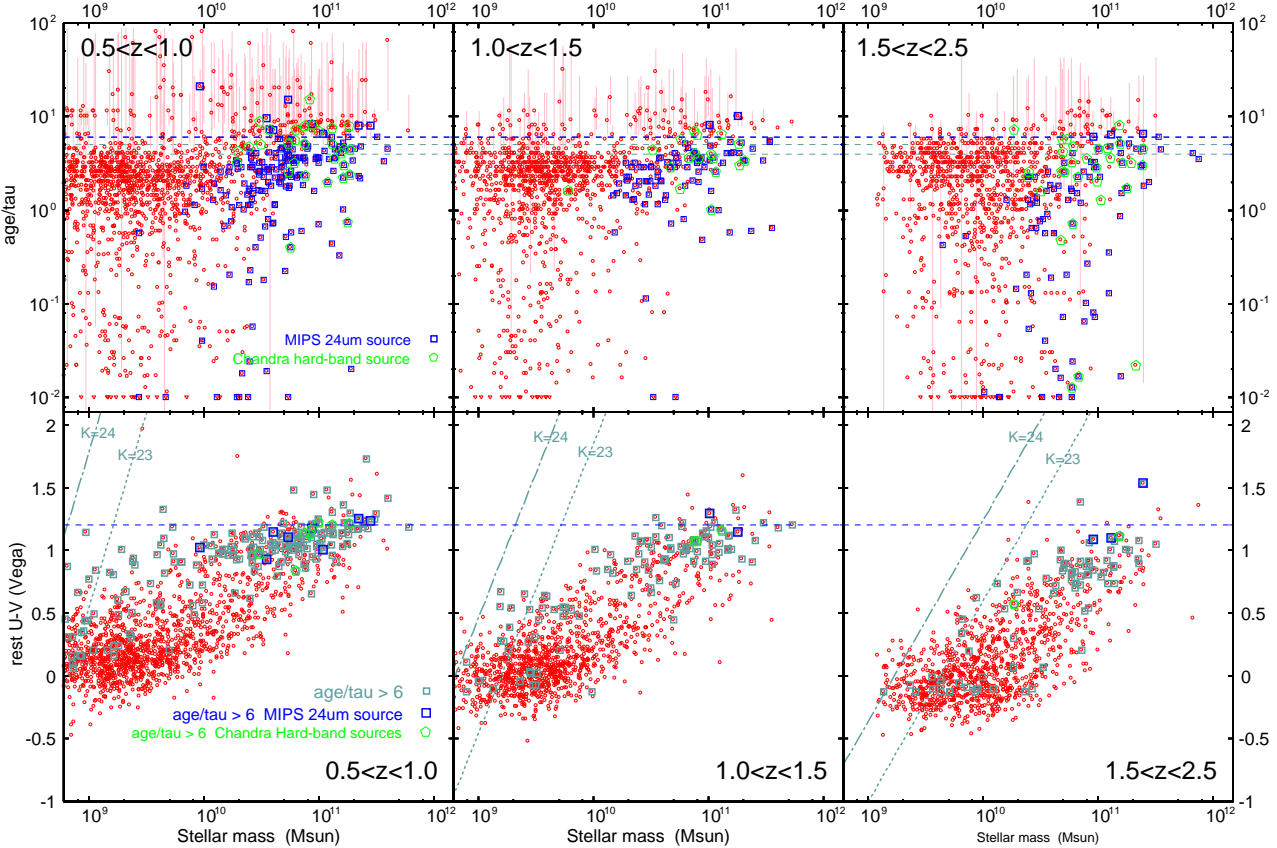}
  \end{center}
\vspace{-4mm}
  \caption{{\bf top:} age/$\tau$ vs. stellar mass for  $K_{s}$-selected galaxies in 
the MODS field in each redshift bin. 
The uncertainty of age/$\tau$ is shown as errorbars for galaxies with 
age/$\tau > 6$. 
Errors in age/$\tau$ also 
include the photometric redshift error for those 
with no spectroscopic identification. 
Blue open squares represent objects detected in the MIPS 24$\mu$m image with  
S/N $>$ 5 (f$_{\rm 24\mu m} \gtrsim 20$ $\mu$Jy). 
The MIPS 24$\mu$m sources with age/$\tau > 6$ are shown as large blue squares.
Pentagons show Chandra hard-band 
sources from the CDF-North catalog \citep{ale03}. 
Thick dashed line shows the criterion of age/$\tau = 6$ for quiescent galaxies 
and thin dashed lines represent age/$\tau = 5$ and 4. 
{\bf bottom:} rest-frame $U-V$ color vs. stellar mass for the same $K_{s}$-selected 
galaxies. Gray open squares show galaxies with age/$\tau > 6$ in the top panels. 
Blue larger squares and green pentagons shows the MIPS 24$\mu$m and Chandra 
hard-band sources with age/$\tau > 6$, respectively.
Horizontal dashed line represents rest $U-V = 1.2$.   
Short-dashed and dashed-dotted lines represent the $K_{s}$-band magnitude limit 
for the wide ($K_{s} = 23$) and deep ($K_{s} = 24$) fields, respectively (see text 
for details).
}
\label{fig:mstau}
\end{figure*}

\section{Sample selection}
\subsection{Selection of quiescent galaxies}

Taking advantages of the deep MOIRCS $JHK_{s}$-bands photometry and 
the wide wavelength coverage of the multi-wavelength ancillary 
data of the GOODS,  we used the results of the SED fitting described in the 
previous section to select quiescent galaxies. 
Several studies have used the SED fitting technique for the selection of 
quiescent galaxies (\cite{zuc06}; \cite{arn07}; \cite{gra07}; \cite{sal08}; \cite{fon09}).
Following \citet{fon09} and \citet{gra07}, 
we use the criterion with age/$\tau$, assuming SFR $\propto \exp(-age/\tau)$ in 
the SED fitting with the GALAXEV model.
In this study, 
we adopt age/$\tau > 6$ as in \citet{fon09}, which roughly corresponds to 
SFR/M$_{\rm star} \lesssim 10^{-11}$ yr$^{-1}$ for relatively small $\tau$ values 
\citep{fon09}.
It should be noted that our lower limit of $\tau \ge $ 0.1 Gyr mentioned in the previous 
section leads to a lower limit 
of age $ >$ 0.6 Gyr for galaxies with age/$\tau > 6$. 
The upper panels of Figure \ref{fig:mstau} 
show the distribution of age/$\tau$ as a function of stellar mass for galaxies at 
$0.5<z<1.0$, $1.0<z<1.5$, and $1.5<z<2.5$.  
Thick dashed line shows the criterion of age/$\tau = 6$ and thin dashed lines 
represent age/$\tau = 5$ and 4.
The MIPS 24$\mu$m sources ($\gtrsim 20 \mu$Jy) are also shown 
in the figure.  
The mid-infrared detection indicates dust enshrouded star formation 
(or AGN activity). 
It is seen that at age/$\tau \sim $ 4--5,  
many galaxies are detected on the MIPS 24$\mu$m image, 
especially at M$_{\rm star} \gtrsim 10^{10.5}$ M$_{\odot}$, while 
there are only a small number of 24$\mu$m sources at age/$\tau >6$.  
The criterion of age/$\tau >$ 5 or 4 could lead to the contamination of 
the quiescent sample with objects with some star-forming activities. 
Therefore we use the criterion of age/$\tau > 6$ in this paper, although 
we will check the results with the criteria of age/$\tau >$ 5 and 4 in Section 
\ref{sec:comp}. For the same reason, we excluded the MIPS 24$\mu$m sources with 
age/$\tau > 6$ from the quiescent sample as in \citet{fon09}. 
There are 9, 2, and 3 such 24$\mu$m sources with age/$\tau > 6$ at 
$0.5<z<1.0$, $1.0<z<1.5$, and $1.5<z<2.5$, respectively.
The 24$\mu$m fluxes of these objects could be powered by AGN 
(e.g., \cite{nar08}). 
We show hard X-ray sources from the Chandra Deep Field North catalog 
\citep{ale03} in Figure 1, and 
only one of these 24$\mu$m sources with age/$\tau > 6$ 
is significantly detected in the hard X-ray data. 
We also 
found that the other 13 24$\mu$m sources have the 1.6$\mu$m bump 
in their near- to mid-infrared SEDs and do not show power-law SEDs. 
Therefore most of 24$\mu$m sources with age/$\tau > 6$ are considered to be 
star-forming galaxies, and 
we excluded all these 24$\mu$m sources from the quiescent sample 
as objects with relatively active star formation.

\begin{figure*}[t]
  \begin{center}
    \FigureFile(170mm,130mm){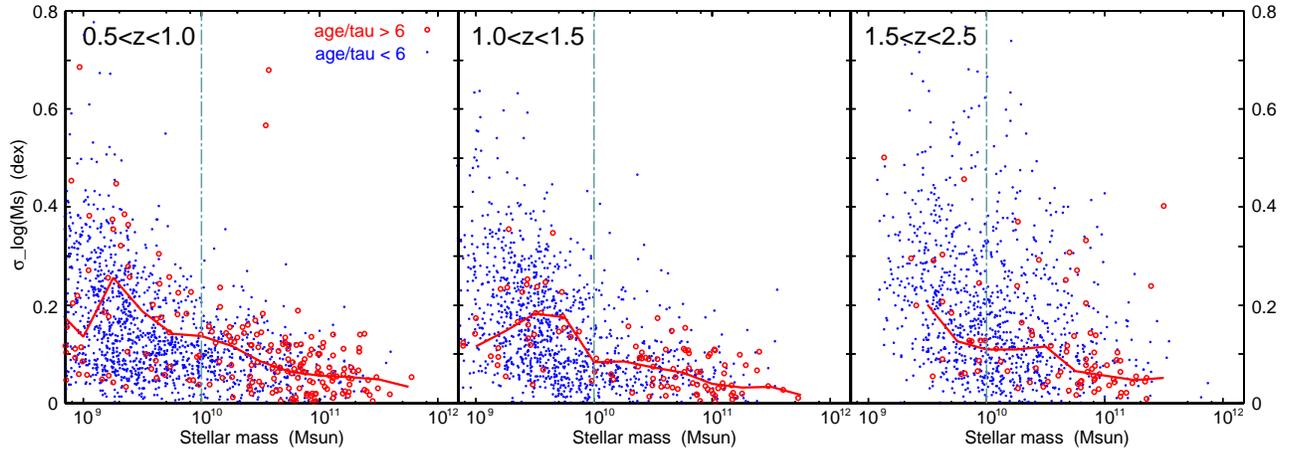}
  \end{center}
\vspace{-4mm}
  \caption{Uncertainty of the estimated stellar mass as a function of stellar mass in 
each redshift bin. Red circles show quiescent galaxies with age/$\tau > 6$ and 
blue dots represent those with age/$\tau < 6$. Solid line shows 
the median values at each stellar mass for quiescent galaxies.
Vertical dashed-dotted line represents the limiting stellar mass for the quiescent sample. 
For objects without spectroscopic redshift, the photometric redshift error is taken into 
account in the estimate of the stellar mass uncertainty. 
 }\label{fig:mserr}
\end{figure*}

\subsection{Limiting stellar mass for quiescent galaxies}
The $K_{s}$-band magnitude-limited sample does not have a sharp limit in 
stellar mass even at a fixed redshift, because the stellar M/L ratio 
at the observed $K_{s}$ band varies with different stellar populations 
(e.g., \cite{kaj06b}; K09).
K09 investigated how the $K_{s}$-band magnitude limit affects the stellar 
mass distribution of the sample by 
using the distribution of the rest-frame $U-V$ color, which reflects 
the stellar M/L ratio well, as a function of stellar mass (Figure 3 in K09).
We here use the similar method to determine the limiting stellar mass for 
quiescent galaxies.
The bottom panels of Figure \ref{fig:mstau} show the rest-frame $U-V$ color 
vs. stellar mass for the $K_{s}$-selected galaxies. 
Quiescent galaxies selected by age/$\tau > 6$ in the previous section 
are shown as gray squares.
The $K_{s}$-band magnitude limits 
for the wide ($K_{s} = 23$) and deep ($K_{s} = 24$) fields are shown as
Short-dashed and dashed-dotted lines, respectively.
All objects with stellar mass larger than the line at a fixed $U-V$ color 
(on the right side of the line in the figure) are brighter than the magnitude limit.
We used the GALAXEV model with various star formation histories to calculate
the line (see K09 for details). 
The slope of the lines represents that 
galaxies with redder $U-V$ colors tend to have higher stellar M/L ratios and 
that at a fixed $K_{s}$-band flux (luminosity), the limiting stellar mass 
for red galaxies is systematically higher than that for blue galaxies. 

In the figure, quiescent galaxies with age/$\tau > 6$ tend to have red 
$U-V$ colors, typically $U-V \gtrsim 0.8$, especially at M$_{\rm star} \gtrsim 10^{10}$ 
M$_{\odot}$, and form the well-known red sequence. 
It is seen that 
the color distribution of the quiescent galaxies shifts to bluer with redshift.
If we select galaxies with $U-V > 0.8$, $U-V > 0.7$, and $U-V > 0.6$ as the 
red sequence population at $0.5<z<1.0$, $1.0<z<1.5$, and $1.5<z<2.5$, 
respectively, following \citet{whi10} and \citet{bra09}, the fraction of quiescent 
galaxies in the red sequence population is 46\% (55\% in the red sequence galaxies with 
M$_{\rm star} > 10^{11}$ M$_{\odot}$), 36\% (40\%), and 32\% (32\%) for the redshift 
bins.
Since most quiescent galaxies show the rest $U-V \sim 1.2$ or bluer, 
we adopt $U-V = 1.2$ (horizontal dashed line in the figure) as a reference  
to determine the limiting stellar mass for the galaxies 
in  the wide and deep fields. At $1.5<z<2.5$, we can nearly completely 
sample the galaxies 
down to $\sim 10^{10.5}$ M$_{\odot}$ for the wide field and to $\sim 10^{10}$ M$_{\odot}$ 
for the deep field. In the following, 
we use quiescent galaxies 
with M$_{\rm star} > 10^{10.5}$ M$_{\odot}$ 
in the wide field and those with M$_{\rm star} > 10^{10}$ M$_{\odot}$ in the deep field 
as a mass-limited quiescent sample at $1.5<z<2.5$. 
For the $0.5<z<1.0$ and $1.0<z<1.5$ redshift bins, 
we also limit the quiescent sample to 
those with  M$_{\rm star} > 10^{10}$ M$_{\odot}$ for a fair comparison with 
the $1.5<z<2.5$ bin, while the limiting stellar mass  
is lower than $10^{10}$ M$_{\odot}$ in the both fields 
for these redshift bins. 
The number of galaxies in the mass-limited quiescent sample is   
130 at $0.5<z<1.0$, 70 at $1.0<z<1.5$ and 50 at $1.5<z<2.5$.

It should be noted that our multi-band data other than $K_{s}$-band are also 
deep enough to select quiescent galaxies down to the limiting stellar mass 
mentioned above.  
For example, even quiescent objects at $z\sim$ 2--2.5 
with M$_{\rm star} > 10^{10}$ M$_{\odot}$ ($> 10^{10.5}$ M$_{\odot}$ for the wide field) 
 whose stellar age is  
 as old as the age of the universe are expected to be detected at least in 
$H$, 3.6$\mu$m, and 4.5$\mu$m bands at S/N $>$ 2 in addition to the 
$K_{s}$-band detection (also in $J$ band for $z\lesssim$ 2.3). 
Furthermore, deep $HST$/ACS data are deep enough to detect 
at least quiescent objects with M$_{\rm star} = 10^{10}$ M$_{\odot}$ 
and with stellar age of $\sim$ 0.6--1.0 Gyr in $V$, $i$, and 
$z$ bands at $z<2.5$. 
Therefore we can judge whether the criteria for the quiescent population are 
satisfied or not for galaxies with M$_{\rm star} \sim 10^{10}$ M$_{\odot}$ 
even at $z\sim$ 2--2.5. 
For dusty galaxies, however, it could not be the case. 
Objects with large extinction are often not detected in the ACS images, and 
the uncertainty in age/$\tau$ of these objects tend to be large. 
In fact, 
galaxies with large errorbars in the upper panels of Figure \ref{fig:mstau} 
have relatively red SEDs in the NIR--MIR region and 
are fitted with dusty model templates. 
Since the number of such objects in the quiescent sample is small, however, 
the contamination from heavily obscured star-forming galaxies does not seem 
to affect our results in the following. 

Figure \ref{fig:mserr} shows the uncertainty of the estimated stellar mass for 
the quiescent galaxies. At M$_{\rm star} \gtrsim 10^{10}$ M$_{\odot}$, the stellar 
mass errors are relatively small and typically $< 0.2$ dex.  
Furthermore, as seen in the next section, the low-mass slope of the stellar mass 
function for quiescent galaxies are nearly flat or slightly positive ($\alpha \sim $ 0--0.6). 
Since it suggests that the number density of low-mass quiescent galaxies is relatively 
small, the contamination from lower-mass galaxies could not be significant even if 
these low-mass galaxies have slightly larger stellar mass errors. 
Therefore the incompleteness and contamination due to the stellar mass error 
are expected to be negligible.

\section{Results}
\label{sec:result}
\begin{figure*}[t]
  \begin{center}
    \FigureFile(170mm,120mm){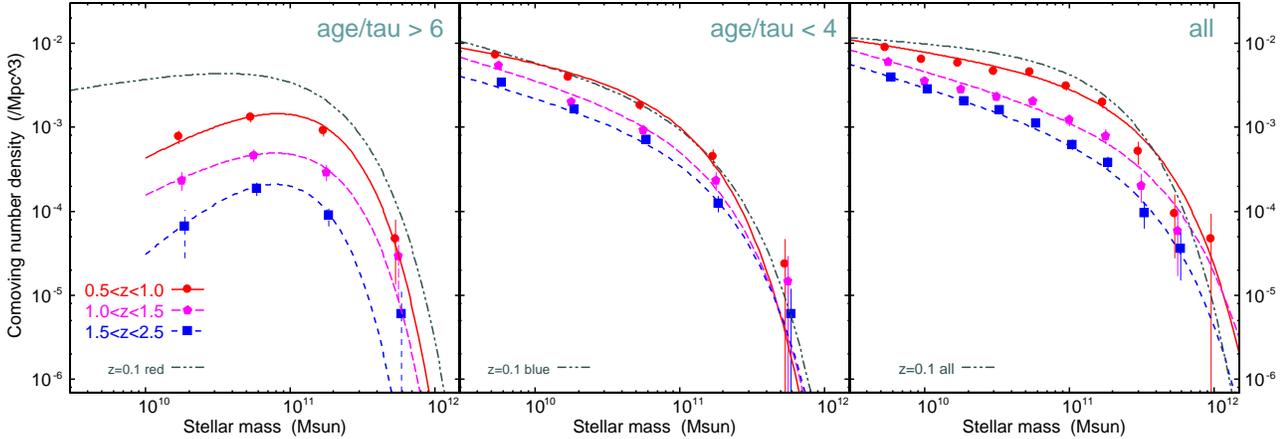}
  \end{center}
\vspace{-4mm}
  \caption{The evolution of the stellar mass function of 
quiescent galaxies with age/$\tau > 6$ 
(left), star-forming galaxies with age/$\tau < 4$ (middle), and  
all stellar mass-selected galaxies (right, from K09). 
Circles, pentagons, and squares show the results calculated with the 
1/V$_{\rm max}$ formalism for galaxies at
 $0.5<z<1.0$, $1.0<z<1.5$, and $1.5<z<2.5$, respectively. 
Error bars are based on the Poisson statistics.
Solid, long-dashed, and short-dashed lines represent the results with the STY method for 
galaxies at $0.5<z<1.0$, $1.0<z<1.5$, and $1.5<z<2.5$, respectively.
For reference, the local SMF for all, red, and blue 
galaxies at $z\sim0.1$ by \citet{bel03} is shown as the double-dotted dashed line in 
each panel.
 }\label{fig:mf}
\end{figure*}

The left panel of Figure \ref{fig:mf} shows the stellar mass function (hereafter, SMF) of 
quiescent galaxies selected in the previous section in the different redshift bins. 
For comparison, 
those for star-forming galaxies with age/$\tau < 4$ and 
all stellar mass-selected galaxies 
are also shown.
We use the V$_{\rm max}$ method to calculate the comoving number density.
K09 calculated the V$_{\rm max}$ for the $K_{s}$-band magnitude limit 
($K_{s}=23$ for the wide field or $K_{s}=24$ for the deep field) 
by using  the best-fit model template for each object (see Section 3.3 in K09 
for details). 

It is seen that the number density of quiescent galaxies gradually 
increases from $1.5<z<2.5$ to $0.5<z<1.0$ in all mass ranges. 
The increase of the number density seems to be nearly independent of 
stellar mass over $10^{10}$--$10^{11.5}$ M$_{\odot}$. 
The number density increases by a factor of $\sim 3$ from $1.0<z<1.5$ to 
$0.5<z<1.0$ and by a factor of $\sim 10$ from $1.5<z<2.5$ to $0.5<z<1.0$ 
in this mass range. 
On the other hand,  the number density of 
star-forming galaxies with age/$\tau < 4$ increases only by a factor of $\sim 1.9$ 
from $1.0<z<1.5$ to $0.5<z<1.0$ and by a factor of $\sim 2.8$ from 
$1.5<z<2.5$ to $0.5<z<1.0$. 
The number density evolution of quiescent galaxies is clearly stronger than 
that of star-forming galaxies. 

\begin{figure*}
  \begin{center}
    \FigureFile(160mm,120mm){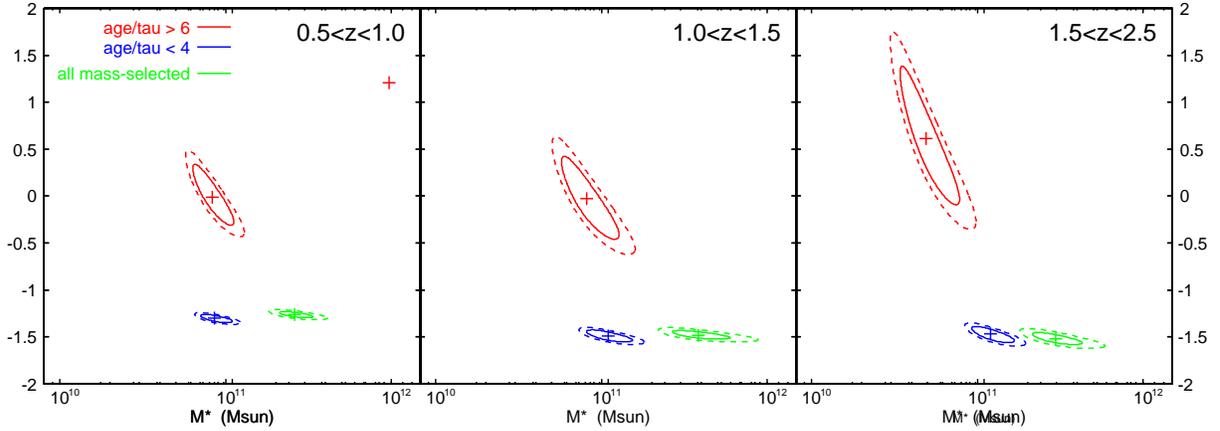}
  \end{center}
\vspace{-4mm}
  \caption{The Schechter parameters in M$^*$--$\alpha$ plane for the quiescent (red),  
star-forming (blue), and all stellar mass-selected (green) samples in each redshift bin.
Crosses show the best-fit values determined with the STY method.
1$\sigma$ (solid) and 2$\sigma$ (dashed) error contours are also shown.
 }\label{fig:malpha}
\end{figure*}

In order to evaluate the shape of the SMF, we used the 
STY method \citep{san79} assuming the Schechter function form \citep{sch76} 
as in K09. We estimated the best-fit values of the Schechter parameters, 
namely, the characteristic mass M$^{*}$ and the low-mass slope $\alpha$ for 
both the quiescent sample and the star-forming galaxies. 
For the star-forming galaxies, we used the same limiting stellar mass as a function 
of redshift as in K09, which is estimated by using  the 90 percentile
 of $U-V$ color at each stellar mass for all stellar mass-selected galaxies 
(see Section 3.2 of K09 for details). Such limiting stellar mass is probably suitable 
(or conservative) for the star-forming galaxies, because these galaxies 
tend to have relatively low stellar M/L ratios except for extremely dusty galaxies. 
Further details of estimating the Schechter parameters with the STY method are 
explained in Section 3.3 of K09.

Figure \ref{fig:malpha}  shows the best-fit Schechter parameters and their 
uncertainty in the M$^*$-$\alpha$ plane for each redshift bin.
 The best-fit values and their uncertainty are also summarized in Table \ref{tab:sche}.
For comparison, we also show the best-fit M$^*$ and $\alpha$ for 
all stellar mass-selected galaxies estimated by K09. 
The low-mass slope $\alpha$ for the quiescent sample is significantly flatter 
than those for the star-forming and all stellar mass-selected samples over 
$0.5<z<2.5$,   
although the uncertainty of $\alpha$ for these quiescent galaxies is 
relatively large due to 
the small number of the sample and the relatively high limiting stellar mass.  
\vspace{1mm}
\begin{table*}[t]
  \caption{Best-fit M$^{*}$ and $\alpha$ obtained with the STY method for quiescent 
and star-forming galaxies}
\label{tab:sche}
\vspace{-3mm}
\begin{center}
    \begin{tabular}{ccccccc}  
   \hline 
\vspace{-3mm}\\
\vspace{1mm}
   & \multicolumn{2}{c}{quiescent (age/$\tau>6$)} & \multicolumn{2}{c}{star-forming (age/$\tau<4$)} & \multicolumn{2}{c}{all mass-selected (from K09)}\\
\vspace{1mm}
   redshift & $\log_{10}$M$^*$(M$_{\odot}$)  & $\alpha$ & $\log_{10}$M$^*$(M$_{\odot}$)  & $\alpha$ & $\log_{10}$M$^*$(M$_{\odot}$)  & $\alpha$\\
\hline 
\vspace{-3mm}\\
\vspace{1mm}
 $0.5<z<1.0$ & 10.92$^{+0.08}_{-0.08}$ & -0.01$^{+0.23}_{-0.19}$ & 10.93$^{+0.07}_{-0.05}$ & -1.30$^{+0.02}_{-0.03}$ & 11.33$^{+0.10}_{-0.07}$ & -1.26$^{+0.03}_{-0.03}$\\ \vspace{1mm}
 $1.0<z<1.5$ & 10.91$^{+0.12}_{-0.10}$ & -0.03$^{+0.29}_{-0.29}$ & 11.04$^{+0.09}_{-0.09}$ & -1.49$^{+0.04}_{-0.04}$ & 11.48$^{+0.16}_{-0.13}$ & -1.48$^{+0.04}_{-0.04}$\\ \vspace{1mm}
 $1.5<z<2.5$ & 10.69$^{+0.12}_{-0.10}$ & +0.61$^{+0.49}_{-0.47}$ & 11.08$^{+0.09}_{-0.07}$ & -1.47$^{+0.05}_{-0.06}$ & 11.38$^{+0.14}_{-0.12}$ & -1.52$^{+0.06}_{-0.06}$\\
       \hline
    \end{tabular}
  \end{center}
\end{table*}
On the other hand, 
the characteristic mass M$^*$ for the quiescent sample is similar with 
that for the star-forming galaxies, while it could be slightly smaller than 
the star-forming galaxies at $1.5<z<2.5$. 
The M$^*$ for the both quiescent and star-forming samples is lower 
than that for all mass-selected galaxies by a factor of $\sim 2$. 
This may be due to the effects of the combination of the 
(quiescent and star-forming) mass functions 
with different shapes (different $\alpha$). 
K09 found a marginal upturn around $\sim 10^{10}$ M$_{\odot}$ in the stellar 
mass function for the stellar mass-selected sample, which cannot be fitted very 
well with the single Schechter function form. 
On the other hand, the SMF for the star-forming galaxies 
seems to have a very weak or no upturn, and is fitted with the Schechter function 
 better than that for all mass-selected galaxies in all redshift bins. 
The contribution of the flatter mass function of quiescent galaxies at 
M$_{\rm star} \gtrsim 10^{10}$ M$_{\odot}$ could cause 
the upturn in the SMF for all mass-selected galaxies, 
which has been observed at $z<1$ (e.g., \cite{dro09}; \cite{pen10}).
Fitting the mass function with such a upturn with the single Schechter function 
may lead to the slightly higher M$^{*}$.  

We found no significant evolution of the shape of the SMF for 
the quiescent sample. 
Although M$^*$ and $\alpha$ for these galaxies might become slightly 
lower and more positive values respectively at $1.5<z<2.5$, the large uncertainty 
prevents us from confirming this. 
The evolution of the shape of the SMF for the star-forming galaxies is similar with 
that for all mass-selected galaxies; the low-mass slope becomes slightly steeper 
with redshift at $z>1$, while the characteristic mass does not evolve significantly.

\begin{figure}[t]
  \begin{center}
    \FigureFile(80mm,100mm){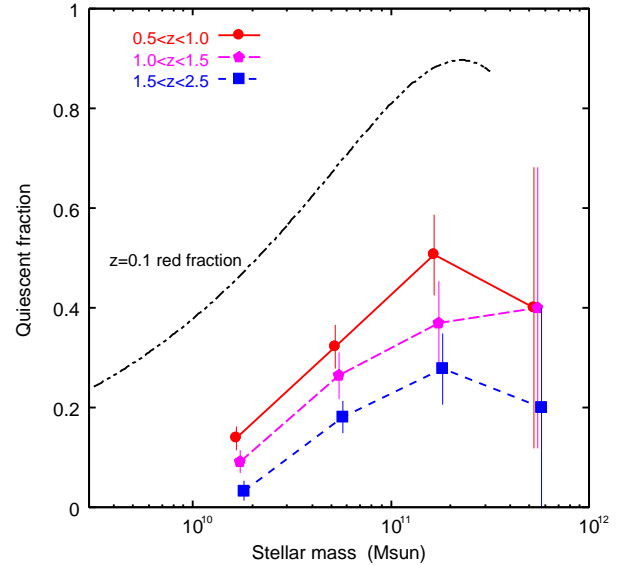}
  \end{center}
\vspace{-4mm}
  \caption{Fraction of quiescent galaxies as a function of stellar mass.
Circles, pentagons, and squares represent galaxies at $0.5<z<1.0$, $1.0<z<1.5$, 
and $1.5<z<2.5$, respectively.
Error bars are based on the Poisson statistics.
For reference, the red fraction for  
galaxies at $z\sim0.1$ from \citet{bel03} is shown as the double-dotted dashed line.
 }\label{fig:pfrac}
\end{figure}

Figure \ref{fig:pfrac} shows the fraction of quiescent galaxies at each stellar mass 
for the different redshift bins. 
The fraction of quiescent galaxies increases with stellar mass over $10^{10}$--$10^{11.5}$ 
M$_{\odot}$ in all redshift bins, though the uncertainty of the fraction at $>10^{11.5}$ 
M$_{\odot}$ is very large due to the very small number of such massive quiescent galaxies
in our sample (1--2 such galaxies in each bin). 
This mass dependence of the fraction of quiescent galaxies over $0.5<z<2.5$ 
is consistent with the above result that the shape of the SMF for 
these galaxies is flatter than those for the star-forming galaxies. 
On the other hand, 
the fraction of quiescent galaxies at each mass decreases with redshift 
 (increases with time) over $10^{10}$--$10^{11.5}$ M$_{\odot}$.   
The quiescent fraction of galaxies with M$_{\rm star} = 10^{11}$--$10^{11.5}$ M$_{\odot}$ 
increases from $\sim$ 25\% at $1.5<z<2.5$ to $\sim$ 50\% at $0.5<z<1.0$, while 
that of galaxies with M$_{\rm star} = 10^{10}$--$10^{10.5}$ M$_{\odot}$ increases from 
$\lesssim$ 5\% to $\sim$ 15\% in the same redshift range.  
This is also consistent 
with the stronger evolution in the number density of quiescent galaxies than 
that of the star-forming galaxies mentioned above. 

\section{Discussion}
\subsection{Comparison with other studies}
\label{sec:comp}

\begin{figure*}[t]
  \begin{center}
    \FigureFile(170mm,120mm){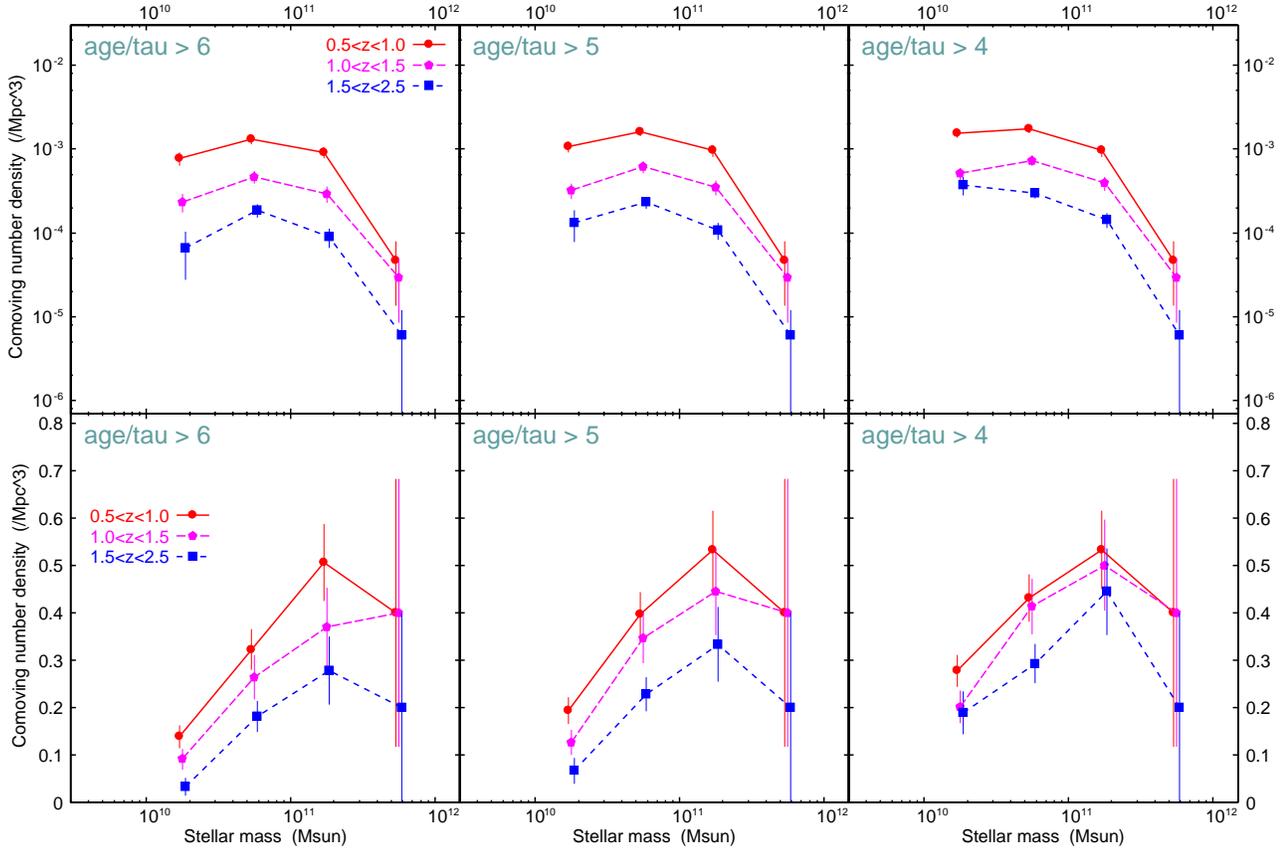}
  \end{center}
\vspace{-4mm}
  \caption{Number density and fraction of quiescent galaxies as a function of 
stellar mass for the different criteria of age/$\tau > 6$ (left), age/$\tau > 5$ (middle), and 
age/$\tau > 4$
 (right). Symbols are the same as Figure \ref{fig:pfrac}. 

 }\label{fig:ptest}
\end{figure*}

We have investigated the evolution of the number density and the fraction of 
quiescent galaxies as a function of stellar mass at $0.5<z<2.5$, using the very 
deep NIR data of the MODS and the multi-wavelength ancillary data of the GOODS. 
Here we compare our results with previous studies in other general fields.

\citet{ver08} investigated the evolution of early-type galaxies selected by spectral 
classification with the $D_{\rm n}$4000 index up to $z\sim1.3$, 
using the VVDS spectroscopic sample ($I_{\rm AB}<24$).
They found that the low-mass slope of the SMF for early-type 
galaxies is flatter ($\Delta\alpha \sim 1$) than those of star-forming and overall 
galaxy population. The number density of early-type galaxies increases by 
a factor of $\sim$ 3 between $1.0<z<1.3$ and $0.7<z<1.0$, while that of 
late-type galaxies increases by a factor of $\sim$ 2 in the same redshift range.
Our results seen in the $0.5<z<1.0$ and $1.0<z<1.5$ bins are consistent with 
these results of \citet{ver08}. 

\citet{ilb10} also studied the evolution of the SMF of quiescent galaxies 
selected by a color selection with the rest-frame $NUV - r'$ and $r' - J$ up to $z\sim2$, 
using the Spitzer/IRAC 3.6$\mu$m-selected sample ($m_{\rm AB}<23.9$) in the 
COSMOS 2 deg$^2$ field. Although their sample reach only to $\sim 10^{11}$ M$_{\odot}$ 
at $z\gtrsim1.5$, they estimated the low-mass slope of the SMF for 
quiescent galaxies up to $z\sim1.2$, and found that the low-mass slope for these galaxies
gradually  flatten with redshift and becomes $\alpha \sim 0$ at $z\sim0.8$. 
The number density of these quiescent galaxies increases by a factor of $\sim 3$ 
from $z\sim1.2$ to $z\sim0.7$ and by a factor of $\sim 10$ from $z\sim2$ to $z\sim0.7$.
Such evolution of quiescent galaxies is consistent with our results seen in the 
previous section. We found that the flat low-mass slope for these galaxies continues  
up to $z\sim2$. \citet{ilb10} also pointed out that the evolution of the SMF
for star-forming galaxies over $0.2<z<2$ is relatively weak and that the number density 
of star-forming galaxies increases by a factor of $\sim 2.5$ between $z\sim2$ and 
$z\sim0.9$. The star-forming galaxies with age/$\tau < 4$ in our sample also show  
the similar evolution (the middle panel of Figure \ref{fig:mf}). 

\citet{fon09} investigated the evolution of the fraction of ``red and dead'' 
galaxies with 
age/$\tau > 6$ among massive galaxies with M$_{\rm star}  > 7 \times 10^{10}$ 
M$_{\odot}$ up to $z\sim3.5$, using the GOODS-MUSIC sample in the GOODS-South 
field. 
Since the survey field layout, the filter set used in the SED fitting, and 
the sample selection in this study are similar with those in \citet{fon09}, 
a direct comparison is possible. 
They found that the fraction of the ``red and dead'' galaxies 
gradually decreases with redshift, and Figure 5 in \citet{fon09} shows that the fraction
 is $\sim$ 50\% at $z\sim0.7$, $\sim$ 40\% at $z\sim1.2$, and $\sim$ 20\% at $z\sim2$.
If we limit our quiescent sample to those with M$_{\rm star} > 7 \times 10^{10}$ 
M$_{\odot}$ as in \citet{fon09}, 
the fraction of quiescent galaxies becomes 52\% at $0.5<z<1.0$, 
37\% at $1.0<z<1.5$, and 29\% at $1.5<z<2.5$.    
We estimated the cosmic variance for these galaxies with the method introduced by
\citet{mos10}, using the clustering strength of $r_{0} \sim $ 10 Mpc for 
quiescent galaxies at $1 \lesssim z \lesssim 2$ observed by \citet{wil09} and
 \citet{har10}.  
The correlation length of $r_{0} \sim $ 10 Mpc corresponds to the galaxy bias 
of $b_{g} \sim $ 2 at $z \sim 0.75$, $b_{g} \sim $ 3 at $z \sim 1.25$, 
and $b_{g} \sim $ 4.5 at $z\sim2$. 
The resulting relative cosmic variance for our quiescent sample is 
$\sigma_{v} \sim $ 0.24 at $0.5<z<1.0$, $\sigma_{v} \sim $ 0.30 at $1.0<z<1.5$, 
 and $\sigma_{v} \sim $ 0.26 at $1.5<z<2.5$. 
If we take account of the cosmic variance and Poisson errors, 
the quiescent fractions in this study and \citet{fon09} agree well within the 
uncertainty.  

On the other hand, \citet{gra07} presented the SMF for 
galaxies with age/$\tau >4 $ and those with age/$\tau < 4$ at $1.4<z<2.5$. 
In order to compare our results directly with those in \citet{gra07} and to 
check the effect of changing the selection criterion to our results, 
we estimated the SMF and the fraction of galaxies with age/$\tau > 4$ 
and 5 as a function of stellar mass, and compared them with those of galaxies with 
age/$\tau > 6$ in Figure \ref{fig:ptest}.  
Although the overall trend does not change significantly when the criteria of 
age/$\tau > 5$ and 4 are used, the low-mass slope becomes slightly steeper and 
the evolution of the number density becomes weaker as the selection 
threshold is loosened. This may be due to the gradual increase of the 
contamination from galaxies with star formation activities. 
For example, 
the fraction of massive quiescent galaxies with M$_{\rm star} = 10^{11}$--$10^{11.5}$ 
M$_{\odot}$ already reaches to $\sim$ 50\% at $z\sim2$ in the case with age/$\tau > 4$. 
This is consistent with the similar number densities 
of galaxies with age/$\tau > 4$ and age/$\tau < 4$ at M$_{\rm star} \sim 10^{11}$ 
M$_{\odot}$ seen in Figure 8 of \citet{gra07}.

In summary, the shape of the SMF of quiescent 
galaxies at $z\sim1$ and the number density evolution of massive quiescent 
galaxies with M$_{\rm star} \gtrsim 10^{11}$ M$_{\odot}$ at $1\lesssim z \lesssim 2$ 
seen in the previous section are consistent with the previous studies in other general 
fields. 
Thus, the field-to-field variance does not seem to significantly affect our results.
Furthermore, in this paper, 
we found that the evolution of the number density of quiescent 
galaxies with relatively low stellar mass ($\sim 10^{10}$--$10^{11}$ M$_{\odot}$) 
is  similar with that of massive ones, 
and that the flat low-mass slope of quiescent galaxies continues up to $z\sim2$.

\subsection{Quenching of star formation and mass-dependent evolution of galaxies}
 In Section \ref{sec:result}, we found that 
the number density of quiescent galaxies with age/$\tau > 6$ 
increases by a factor of $\sim 3$ from $1.0<z<1.5$ to 
$0.5<z<1.0$ and by a factor of $\sim 10$ from $1.5<z<2.5$ to $0.5<z<1.0$ 
over M$_{\rm star} = 10^{10}$--$10^{11.5}$ M$_{\odot}$. 
We note that 
the number density of galaxies at $0.5<z<1.0$ in our sample could be slightly 
overestimated due to the known large-scale structures around the HDF-North 
(K09; \cite{wir04}).
Nevertheless, 
the rapid increase of the number of 
quiescent galaxies in the universe seems to occur at $1 \lesssim z \lesssim 2$.
Furthermore, 
the evolution of the number density is nearly independent of 
stellar mass. The low-mass slope of the SMF for these 
quiescent galaxies is significantly flatter than those for star-forming and all stellar 
mass-selected galaxies over $0.5<z<2.5$.
The fraction of quiescent galaxies for massive galaxies is higher than 
that for low-mass galaxies in the redshift range. 
Here we consider that the increase of quiescent galaxies is caused by the cessation of 
star formation in some fraction of star-forming galaxies as in previous studies at 
$z<1$ (e.g., \cite{bel07}; \cite{pen10}). 
Since the number density of quiescent galaxies rapidly increases by a factor of $\sim 3$ 
between $z\sim 2$ and $z\sim1.25$ ($\sim 1.6$ Gyr) and by a factor 
of $\sim 3$ between $z\sim 1.25$ and $z\sim 0.75$ ($\sim 2.3$ Gyr), 
the quenching of star formation is expected to occur preferentially in more massive 
galaxies over $1 \lesssim z \lesssim 2$ in order to maintain the mass-dependence of 
the fraction of quiescent galaxies mentioned above. 
We estimated the quenching rate as a function of stellar mass  
by calculating the fraction of 
newly emerging quiescent galaxies between the redshift bins relative to the 
star-forming population including newly increased galaxies at a given stellar mass range.
For simplicity, other processes affecting this fraction such as 
the hierarchical merging is ignored.
As a result, we expect that 
7\%, 18\% and 29\% of the star-forming galaxies with 
M$_{\rm star} = 10^{10}$--$10^{10.5}$ M$_{\odot}$, 
$10^{10.5}$--$10^{11}$ M$_{\odot}$  and $10^{11}$--$10^{11.5}$ 
M$_{\odot}$ ceased star formation between $1.5<z<2.5$ and $1.0<z<1.5$, 
respectively, and 10\%, 23\% and 41\% of the star-forming galaxies with the 
same stellar mass ranges became quiescent 
between $1.0<z<1.5$ and $0.5<z<1.0$. 
It seems that 
more massive star-forming galaxies tend to cease star formation preferentially 
at $1 \lesssim z \lesssim 2$.
Recently, \citet{pen10} reported that such a mass-dependent quenching process 
could explain no evolution of the shape of the SMF for star-forming 
galaxies and the flatter low-mass slope and the similar characteristic 
mass for passive galaxies at $z<1$. 
Such ``mass quenching''  might work even at $1 \lesssim z \lesssim 2$, when 
the rapid increase of the number density of quiescent galaxies was observed. 
The characteristic mass M$^*$ of the SMF for 
the star-forming galaxies with age/$\tau < 4$ 
in our sample shows no significant evolution at the 
redshift range (middle panel of Figure \ref{fig:mf}), 
which seems to be also consistent with the prediction by \citet{pen10}.

K09 found the mass-dependent evolution of the number density
 for all stellar mass-selected 
galaxies at $z\gtrsim $ 1--1.5. 
The number density of galaxies with $M_{\rm star} \sim 10^{11}$ M$_{\odot}$ 
 more strongly evolves than that for galaxies with $M_{\rm star} \sim 10^{10}$ M$_{\odot}$ 
(right panel of Figure \ref{fig:mf}).
On the other hand, the increase of the number density of quiescent galaxies 
is nearly independent of stellar mass as discussed above. 
The mass dependence of the evolution for star-forming galaxies is also 
rather weak (middle panel of Figure \ref{fig:mf}). 
The mass-dependent evolution for all population could be explained  
 by the differences between the quiescent and star-forming populations 
in the strength of the evolution of the number density and in the shape of the
SMF. 
The quiescent population has the flatter low-mass slope of the SMF  
and shows the stronger evolution of the number density than star-forming galaxies.
Therefore the rapid increase of quiescent galaxies contributes strongly to 
the number density evolution for galaxies with $\sim 10^{11}$ M$_{\odot}$, while their 
contribution is negligible for those with $\sim 10^{10}$ M$_{\odot}$ because of the rather 
low fraction of these galaxies.
In this context, the more rapid increase of the number density of $\sim $ M$^*$ 
galaxies than lower-mass galaxies at $z \gtrsim$ 1--1.5 
found in K09 might be explained by 
a mass-dependent quenching mechanism 
which preferentially ceases star formation in these massive galaxies. 

\begin{figure*}[t]
  \begin{center}
    \FigureFile(175mm,125mm){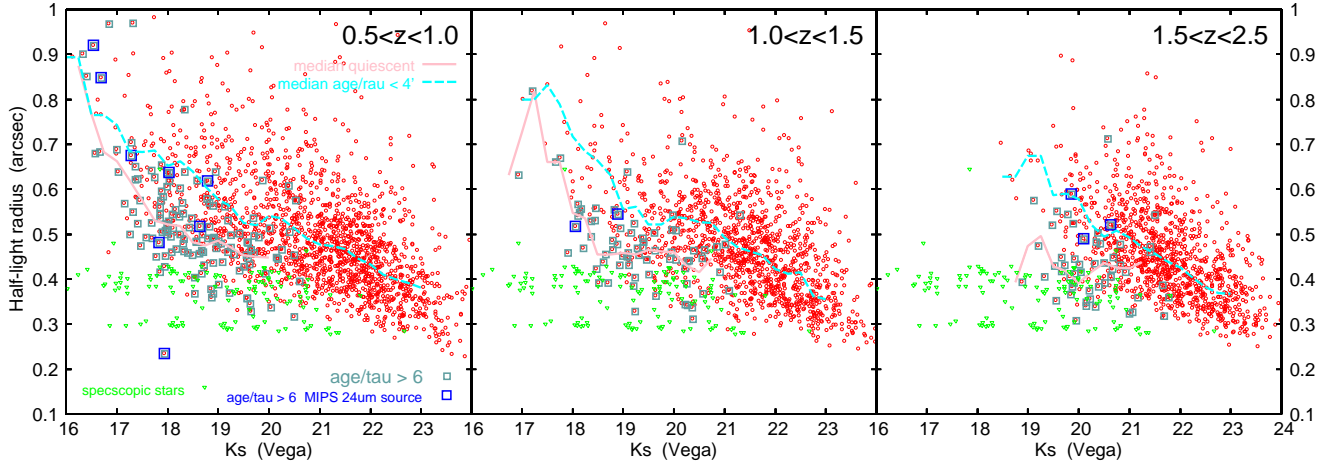}
  \end{center}
\vspace{-4mm}
  \caption{
Half light radius in the observed $K_{s}$ band vs. $K_{s}$-band magnitude for 
$K_{s}$-selected galaxies in each redshift bin. 
Gray squares represent the mass-selected 
(M$_{\rm star} > 10^{10}$ M$_{\odot}$) quiescent galaxies with age/$\tau > 6$. 
Larger blue squares show MIPS 24$\mu$m sources with age/$\tau > 6$. 
Solid and dashed lines represent the median values for the quiescent sample  
and star-forming galaxies with age/$\tau < 4$ at each magnitude, respectively.
It should be noted that the $K_{s}$-band 
half light radii are measured on the mosaic image where  
the PSF varies among the different pointings of 
MOIRCS (FWHM $=$ 0.45--0.60 arcsec,  see \cite{kaj11} for details). 
For reference, spectroscopically confirmed stars are shown as green triangles 
in each panel.
}
\label{fig:krad}
\end{figure*}

In \citet{yam09}, we found that X-ray selected AGNs at $2<z<4$
are preferentially associated with more
massive galaxies, and that $\sim$ 30\% of galaxies with M$_{\rm star} > 10^{11}$
M$_{\odot}$ show the AGN activity with the X-ray luminosity larger than $\sim 10^{42}$
erg s$^{-1}$. \citet{bru09} also found a similar mass-dependence of the
AGN fraction at $1\lesssim z \lesssim 4$.
The AGN feedback may be related with the mass-dependent 
star formation quenching.
\citet{bun08} investigated the fractions of red galaxies and X-ray selected AGNs 
in stellar mass-selected galaxies at $0.4<z<1.4$ 
as a function of stellar mass to compare the star formation 
quenching rate with the AGN triggering rate. They found that 
the mass-dependence and normalization of the both rates agree well, although 
the estimated normalization of the AGN triggering rate 
depends strongly on the assumed AGN lifetime. 
In order to compare our results with the star formation quenching rate in 
\citet{bun08}, we divided the quenching rates calculated above by the time intervals  
between the redshift bins. The results are 4\% Gyr$^{-1}$, 10\% Gyr$^{-1}$, 
and 18\% Gyr$^{-1}$ between $z \sim 1.25$ and $z \sim 0.75$ 
for galaxies with $10^{10}$--$10^{10.5}$ M$_{\odot}$, 
$10^{10.5}$--$10^{11}$ M$_{\odot}$, and $10^{11}$--$10^{11.5}$ M$_{\odot}$, 
respectively, and 4\% Gyr$^{-1}$, 11\% Gyr$^{-1}$, and 18\% Gyr$^{-1}$ between 
$z \sim 2$ and $z \sim 1.25$ for the same mass ranges. 
Both the mass-dependence and normalization are consistent with 
the results of \citet{bun08}, although the selection of quiescent galaxies 
is different between \citet{bun08} (rest-frame $U-B$ color) and this study
(age/$\tau$ from the SED fitting).
Our results may also suggest that the star formation 
quenching rate per unit time 
at a given stellar mass seems to be roughly constant over 
$1 \lesssim z \lesssim 2$. 
In a future work,  
we will study the evolution of the AGN hosts in the MODS field at 
$1 \lesssim z \lesssim 2$ 
and compare it with the star formation quenching 
rate estimated here. 

On the other hand, several studies reported that the clustering of quiescent 
galaxies is stronger than that of star-forming galaxies even at $z\sim2$ 
(\cite{har08}; \cite{har10}; \cite{wil09}), which implies that quiescent galaxies tend to 
inhabit more massive dark matter halos when their number density rapidly increases. 
Therefore the dark matter halo mass 
may play some role in the quenching of star formation. 
\citet{wil09} and \citet{har10} pointed out the possibility that the clustering strength 
of quiescent galaxies at $1 \lesssim z \lesssim 2$ is nearly independent of 
their rest-frame $K$-band luminosity (approximately stellar mass), while 
it is observed at the same redshift range that 
the clustering strength of the overall galaxy population is correlated with stellar 
mass (e.g., \cite{ich07}; \cite{fou10}; \cite{wak10}). 
There may be a critical halo mass above which gas cooling and star formation 
are shut down, for example,  due to the shock heating (e.g., \cite{dek06}; \cite{cat06}). 
In this scenario, the (stellar) mass-dependent quenching of star formation 
 could be explained if massive star-forming galaxies tend to inhabit 
more massive dark matter halos than low-mass star-forming ones, which 
is naturally predicted by theoretical models where the star formation activity 
is simply regulated by the gas (and dark matter) accretion rate in the halo 
(e.g., \cite{bou10}).
In this context, it is interesting to investigate 
the mass-dependence of the clustering strength for quiescent 
and star-forming samples at $1 \lesssim z \lesssim 2$ more intensively.  

The quenching of star formation may be also related with the surface stellar 
mass density of galaxies. Several studies at low and high redshifts  
reported that the rest-frame color or specific star formation rate of galaxies 
are strongly correlated with their surface stellar mass density (e.g., \cite{kau06}; 
\cite{fra08}; \cite{wil10}). 
Most galaxies with the surface mass density higher than a threshold value have 
red color and low star formation activities at each redshift. The threshold surface  
density seems to  increase with redshift gradually. 
At $z\sim2$, many studies have found 
massive quiescent 
galaxies with much smaller sizes (e.g., R$_{\rm e} \sim$ 1 kpc) than 
the present massive ellipticals (\cite{dad05}; \cite{zir07}; \cite{lon07}; 
\cite{cim08}; \cite{van08}; \cite{dam09}; \cite{rya10}).
Quiescent galaxies in our sample also tend to have small half light radii 
in the observed $K_{s}$ band relative to star-forming galaxies with similar 
$K_{s}$-band magnitudes (Figure \ref{fig:krad}). Considering 
their higher stellar M/L ratio and smaller size than star-forming galaxies, 
these quiescent galaxies are expected to have higher surface stellar mass densities. 
At $1.5<z<2.5$, there are also $K_{\rm s}$-bright 
quiescent galaxies whose sizes cannot be distinguished 
from point sources. 
Since compact galaxies with R$_{\rm e} \sim$ 1 kpc at $z\gtrsim 1$ 
 are expected 
not to be resolved in the MODS $K_{s}$-band image, these galaxies may be such 
compact quiescent objects. 
In order to estimate the surface stellar mass density of galaxies quantitatively 
 and investigate 
the relation between the mass-dependent evolution of quiescent galaxies 
and the surface stellar mass density,  
 we need a detailed analysis of the surface brightness profiles which 
 takes account of the seeing effect or higher-resolution 
NIR data such as those with $HST$/WFC3. 

\section{Summary}

We have investigated the evolution of quiescent galaxies at 
$0.5<z<2.5$ in the GOODS-North field as a function of stellar mass, 
using the very deep NIR data of the MODS and 
the public multi-wavelength data of the GOODS.
We performed the SED fitting of the multi broad-band photometry with the GALAXEV 
population synthesis model and selected galaxies with age/$\tau > 6$ and 
without the MIPS 24$\mu$m detection (f$_{\rm 24\mu m} \lesssim 20\mu$Jy) 
as the quiescent population. 
The deep MODS data allow us to construct a stellar mass-limited sample of 
quiescent galaxies down to $\sim 10^{10}$ M$_{\odot}$ even at $z\sim2$ for the 
first time.
Main results in this study are as follows.
\begin{itemize}

\item The number density of quiescent galaxies rapidly increases by a factor of 
$\sim 3$ from $1.0<z<1.5$ to $0.5<z<1.0$ and by a factor of $\sim 10$ from 
$1.5<z<2.5$ to $0.5<z<1.0$. On the other hand, that of star-forming galaxies 
with age/$\tau < 4$ increases only by factors of $\sim 2$ and $\sim 3$ in the 
same redshift ranges.  

\item The increase of the number density of quiescent galaxies is nearly independent 
of stellar mass over $10^{10}$--$10^{11.5}$ M$_{\odot}$, and these galaxies show no 
significant evolution in the shape of the SMF.

\item The low-mass slope of the SMF for these quiescent galaxies 
is $\alpha \sim $ 0--0.6, which is significantly flatter than those of 
the star-forming galaxies and all stellar mass-selected galaxies ($\alpha \sim $
 -1.3 -- -1.5). 
 
\item The fraction of quiescent galaxies increases with stellar mass over $0.5<z<2.5$,
 and decreases with redshift over 10$^{10}$--10$^{11.5}$ M$_{\odot}$; 
the quiescent fraction of galaxies with $10^{11}$--$10^{11.5}$ M$_{\odot}$ 
increases from $\sim $ 25\% at  
$1.5<z<2.5$ to $\sim$ 50\% at $0.5<z<1.0$, while that of galaxies with 
 $10^{10}$--$10^{10.5}$
 M$_{\odot}$ increases from $\lesssim $ 5\% to $\sim $ 15\% in the same redshift range.

\item Provided that the increase of the number density of quiescent galaxies 
is caused by the quenching of star formation in some star-forming galaxies, 
the quenching process is expected to be more effective in more massive galaxies 
in order to maintain the mass-dependence of the fraction of quiescent galaxies 
over $0.5 < z < 2.5$.

\item Since the mass dependence of the evolution of the number density 
 of star-forming galaxies is also weak,  stronger evolution 
of quiescent galaxies with the flatter shape of the SMF 
would cause the more rapid increase of the number density of 
galaxies with M$_{\rm star} \sim 10^{11}$ M$_{\odot}$  
than lower-mass galaxies at $1 \lesssim z \lesssim 2$.

\end{itemize}

\bigskip

We thank an anonymous referee for very helpful suggestions and comments. 
We also thank Yoshi Taniguchi for useful discussions. 
This study is based on data collected at Subaru Telescope, which is operated by
the National Astronomical Observatory of Japan. 
This work is based in part on observations made with the Spitzer Space
Telescope, which is operated by the Jet Propulsion Laboratory,
California Institute of Technology under a contract with NASA.
Some of the data presented in this paper were obtained from the Multi-mission
Archive at the Space Telescope Science Institute (MAST).
STScI is operated by the Association of Universities for Research in
Astronomy, Inc., under NASA contract NAS5-26555.
Support for MAST for non-HST data is provided by the NASA Office of
Space Science via grant NAG5-7584 and by other grants and contracts.
Data  reduction and analysis 
were carried out on common use data analysis computer system 
 at the Astronomy Data Center, ADC, of the National Astronomical 
Observatory of Japan.
IRAF is distributed by the National Optical Astronomy Observatories,
which are operated by the Association of Universities for Research
in Astronomy, Inc., under cooperative agreement with the National
Science Foundation.



\end{document}